\documentclass[10pt,pra,aps,twocolumn,superscriptaddress,floatfix,notitlepage,nofootinbib]{revtex4-1}
\usepackage[latin1]{inputenc}
\usepackage{natbib}
\usepackage{amsmath}
\usepackage{amsfonts}
\usepackage{amssymb}
\usepackage{bm}
\usepackage[pdftex]{graphicx}
\usepackage{graphicx}
\begin{document}
	\title{Black holes as Andreev reflecting mirrors}
	\author{Sreenath K. Manikandan}
	\email{skizhakk@ur.rochester.edu}
	\affiliation{Department of Physics and Astronomy, University of Rochester, Rochester, NY 14627, USA}
	\affiliation{Center for Coherence and Quantum Optics, University of Rochester, Rochester, NY 14627, USA} \author{Andrew N. Jordan}
	\affiliation{Department of Physics and Astronomy, University of Rochester, Rochester, NY 14627, USA}
	\affiliation{Center for Coherence and Quantum Optics, University of Rochester, Rochester, NY 14627, USA}    
	\affiliation{Institute for Quantum Studies, Chapman University, Orange, CA, 92866, USA}
\date{\today}
	\begin{abstract}
We propose a microscopic quantum description for Hawking radiation as Andreev reflections, which resolves the quantum information paradox at black hole event horizons. The detailed microscopic analysis presented here reveals how a black hole, treated as an Andreev reflecting mirror, provides a manifestly unitary description of an evaporating black hole, expanding our previous analysis presented in [PRD 96, 124011 (2017), PRD 98, 124043 (2018)]; In our analogy, a black hole resolves the information paradox by accepting particles -- pairing them with the infalling Hawking quanta into a Bardeen-Cooper-Schrieffer (BCS) like quantum ground state -- while Andreev reflecting the quantum information as encoded in outgoing Hawking radiation. The present approach goes beyond the black hole final state proposal by Horowitz and Maldacena [JHEP 02, 008 (2004)], by providing necessary microscopic details which allows us to circumvent important shortcomings of the black hole final state proposal. We also generalize the present Hamiltonian description to make an analogy to the apparent loss of quantum information possible in an Einstein-Rosen bridge, via crossed Andreev reflections. \end{abstract}
	\maketitle
\section{Introduction}
Andreev reflections are a mode conversion process at the interface between a normal metal and a superconductor, originally discussed by A. F. Andreev to describe the anomalous thermal resistance of a superconductor in the intermediate state~\cite{andreev1964thermal}. It is a special scattering event that involves mode conversions between particle and hole-like modes, exchanging a Cooper pair of electrons with the superconducting condensate~\cite{kummel1969dynamics,mathews1978quasiparticle,blonder1982transition,nagato1993theory,artemenko1978excess,andreev1964thermal,artemenko1979theory,artemenko1979excess,klapwijk2004proximity,dolcini2009andreev,beenakker2000does,courtois1999spectral}. 
Reflecting an incoming mode without changing its momentum is a nontrivial problem, especially in the limit where $\Delta\ll E_{F}$ where the incoming electron has high kinetic energy and the superconducting barrier is weak, yet unable to transmit the electron as there are no allowed electron states within the energy gap $\Delta$. Beenakker describes the process as similar to an ``unmovable rock meeting an irresistible object"~\cite{beenakker2000does}. The superconductor resolves this paradoxical situation by Andreev reflecting a hole-like quasiparticle instead, that has approximately the same momentum as the incoming electron.

Analogous retro-reflection processes from the interface between a normal fluid and superfluid state of bosons have also been discussed~\cite{zapata2009andreev}, which involves exchange of a pair of bosons with the superfluid condensate. Another remarkably  similar problem in solid state physics having some correspondence to Andreev reflections is the Klein tunneling process~\cite{beenakker2008colloquium,beenakker2008correspondence,klein1929reflexion,lee2019perfect};  In the original relativistic situation discussed by Klein, a potential barrier can surprisingly become transparent to incident electrons below the potential, resulting in perfect transmission~\cite{klein1929reflexion}. See Refs.~\cite{calogeracos1999history,dombey1999seventy} for a comprehensive discussion of the problem.

Andreev reflections have found new realms of interest recently as a potential mechanism to resolve major paradoxes pertaining to the quantum description of black holes~\cite{jacobson1996origin,manikandan2017andreev,manikandan2018bosons}. Possible implications for Andreev reflections in black hole thermodynamics was first discussed by Jacobson~\cite{jacobson1996origin} as a resolution to the trans-Planckian reservoir problem, which  in  Hawking's  original calculation appears as the presence of frequencies exceeding the Planck scale~\cite{hawking1975particle}; The frequency of modes propagating just outside the event horizon are redshifted by arbitrary large amounts prior to escaping as outgoing modes -- as a result, the modes associated to the spectrum of frequencies which can be measured by distant observers at later times would have had to originate with very high frequencies, including frequencies exceeding the Planck scale~\cite{parker2009quantum}. One would doubt the validity of quantum field theory at correspondingly high energies. Therefore the question relevant to various semi-classical treatments of black hole evaporation is to describe possible ways in which these outgoing modes can exist, without having to depend on a reservoir of ultrahigh (trans-Planckian) frequencies. 

The trans-Planckian problem has been approached from different directions in the literature (see for instance, Ch. 4.6, ref.~\cite{parker2009quantum}); the work of Unruh, discussing a sonic analogue to the event horizon by considering sound waves propagating in a moving fluid~\cite{unruh1981experimental,unruh1995sonic}, suggested that the ultra-high frequencies appearing in the original work of Hawking~\cite{hawking1975particle,hawking1976breakdown} may not be necessary to obtain the Hawking thermal spectrum in a sonic black hole.     Jacobson's approach in~\cite{jacobson1996origin}, considering Unruh's sonic black hole analogy, suggested that the origin of outgoing Hawking modes at the event horizon can be explained from mode conversion processes similar to Andreev reflections, and therefore not having to  rely on a Trans-Planckian reservoir at the horizon. 

This analogy has been explored further from the perspective of black hole information mirror models that resolve the quantum information paradox~\cite{horowitz2004black,hayden2007black,lloyd2014unitarity} in references~\cite{manikandan2017andreev,manikandan2018bosons}: The analogy maps the interior of a black hole to the superfluid condensate, the exterior to the normal metal/fluid, the interface between normal metal/fluid and the superconductor/superfluid to the event horizon in black holes, and Hawking radiation~\cite{hawking1975particle} to Andreev reflections from the interface.
The information mirror models~\cite{hayden2007black,horowitz2004black} suggest that a black hole, in its late stages of the evaporation process\footnote{The black hole information mirror models~\cite{hayden2007black,horowitz2004black,lloyd2014unitarity} describe an evaporating black hole in its late stages of the evaporation process, where the black hole has radiated away half of its initial entropy (a black hole past the ``half-way point"~\cite{hayden2007black}).}, accepts particles, while reflecting the quantum information in the outgoing modes. In the Horowitz-Maldacena  model~\cite{horowitz2004black}, this is achieved by conjecturing a unique quantum final state at the black hole singularity. Unitarity is ensured when the interactions within the black hole are maximally entangling~\cite{gottesman2004comment,lloyd2014unitarity}, where the model suggests that a black hole in its late stages of the evaporation process can teleport or swap the quantum information by encoding it in the outgoing Hawking radiation. 

Alternatively, the black hole final state proposal~\cite{horowitz2004black} can be viewed as the black hole imposing a special quantum final state boundary condition for the infalling modes. When the final state corresponds to a maximally entangled state, it can act like a fixed point in the Hilbert space, while respecting unitarity of the processes involving the final state. The analogy presented in references~\cite{manikandan2017andreev,manikandan2018bosons} primarily suggested that superfluid quantum ground states of fermions and interacting bosons respectively have several desired qualities to be considered as this final quantum ground state for the modes falling into a black hole. In the analogy, Andreev reflection processes were described as a physical process at the interface between the superfluid and the normal fluid, that preserves quantum information without changing the quantum ground state of the superfluid. Effectively, the superfluid wavefunction acts like a fixed point in the Hilbert space, respecting unitarity of mode conversion processes happening at the interface with a normal fluid.~\footnote{We resort to brief accounts of the analogy in the present article, but we request a careful reader to look at references~\cite{manikandan2018bosons,manikandan2017andreev} where the analogy is developed in detail from both information theoretic and thermodynamic considerations.}

While the description of Andreev reflection as resulting from applying a final state boundary condition is a useful approach to discuss the analogy between Andreev reflections and the quantum physics of a black hole~\cite{manikandan2018bosons,manikandan2017andreev}, we note that resolving the shortcomings of the Horowitz-Maldacena model using this analogy requires one to expand beyond the details of the final state projection approach. One of the major criticisms on the final state projection approach for black hole evaporation is that unitarity of the scattering matrix is assured only when the interactions are maximally entangling, as pointed out by Gottesman and Preskill~\cite{gottesman2004comment}. Small departures from unitarirty in final state projection can lead to superluminal signalling, and computational enhancements beyond that of a standard quantum computer~\cite{almheiri2013black,bao2016grover,lloyd2014unitarity,gottesman2004comment,brun2012perfect,PhysRevD.84.025007,PhysRevLett.106.040403,aaronson2005quantum}, further suggesting inadequacy of the final state projection approach to fully describe the dynamics.    Our present approach resolves these important issues by considering a fully microscopic quantum description of Andreev reflections developed by Nakano and Takayanagi~\cite{nakano1994second}. We discuss connections between the projection approach and the microscopic model at relevant places. Although we only discuss the fermonic case in the present article, we note that a similar analysis should also hold for bosons. 

We emphasize that, albeit the shortcomings, the final state projection approach is indeed an insightful description when the physics is described as a scattering process, where the microscopic details of the scattering center are either inaccessible (for example, in the context of a black hole), or can be ignored. On the other hand, developing analogies as such, to contexts where the microscopic details are readily available, helps us to make an ansatz about the microstates of the inaccessible system, and improve our understanding of its governing dynamical laws. 

This article is organized as follows. We begin with a brief overview of the Horowitz-Maldacena model~\cite{horowitz2004black}. We then proceed to discussing Hawking radiation as Andreev reflections, by adapting the Nakano and Takayanagi description of Andreev reflections~\cite{nakano1994second}. We then show that such a microscopic description allows us to describe the transfer of spin quantum information in Andreev reflections  as a manifestly unitary process, beyond the final state projection approach previously studied~\cite{manikandan2017andreev,manikandan2018bosons}. We point out crucial similarities to the final state projection approach; Andreev reflection proceeds by exchanging an ``information-less" Cooper pair with the black hole final state as discussed previously in the information mirror models~\cite{manikandan2017andreev,manikandan2018bosons}. We comment on how this is equivalent to applying a final state boundary condition (yielding a scattering matrix which is unitary~\cite{horowitz2004black,manikandan2017andreev}), when treated as a scattering process, while differing in microscopic details that overcome shortcomings of the final state projection approach~\cite{gottesman2004comment,almheiri2013black,bao2016grover}.  We also discuss the implications of our results for the quantum physics of black holes and Einstein-Rosen bridges~\cite{einstein1935particle}, and make comparisons to more recent proposals resolving the information paradox, presented in~\cite{piroli2020random,agarwal2019toy,akers2019simple,landsman2019verified,yoshida2017efficient,blok2020quantum}.

\section{The Horowitz-Maldacena black hole final state proposal}
The black hole final state proposal by Horowitz and Maldacena~\cite{horowitz2004black} is an intriguing attempt to resolve the tension between certain string theories~\cite{banks1997m,itzhaki1998supergravity,gubser1998gauge,witten1998anti,maldacena1999large} which suggests that the formation and evaporation of a black hole is a unitary process, and semiclassical descriptions where pure states apparently evolve into mixed states~\cite{hawking1975particle,hawking1976breakdown}. Horowitz and Maldacena suggest adapting an unconventional, but known modification of standard quantum mechanics~\cite{aharonov1964time,aharonov1988result,duck1989sense} to resolve the problem which necessitates a particular quantum final state boundary condition at the black hole singularity. Such a modification circumvents the requirement of tracing over the inaccessible degrees of freedom inside a black hole, and therefore avoids scenarios where pure states can evolve into mixed states. In addition to that, an appropriately chosen unique quantum final state, where the collapsing matter is paired with an infalling Hawking quantum, ensures that quantum information is reflected in the outgoing Hawing quantum, resolving the black hole information paradox~\cite{horowitz2004black,gottesman2004comment,lloyd2014unitarity}.

Horowitz and Maldacena focuses their discussion on evaporating black holes with a space-like curvature singularity, like the Schwarzschild black hole~\cite{schwarzschild1916uber}, and they discuss possible generalizations. Similar to other semiclassical treatments, the Horowitz-Maldacena black hole final state proposal assumes that a local quantum field theoretical description is valid near the event horizon, and one can factorize the Hilbert space across the horizon such that,
\begin{equation}
H=H_{m} \otimes H_{i}\otimes H_{o},
\end{equation}
where $H_{m}, H_{i}~\text{and}~H_{o}$ represent the Hilbert spaces of collapsing matter, states of quantized fluctuations localized inside, and outside the horizon, respectively. In particular, the factorization $H_{i}\otimes H_{o}$ discriminates between the states across the event horizon where properties such as entanglement across the horizon can be defined between an outgoing Hawking mode and a Hawking partner mode trapped inside the horizon (the Unruh state~\cite{unruh1976notes}). The joint Hilbert space, $H_{m}\otimes H_{i}$, describes the ``interior" of a collapsing black hole, including the collapsing matter, but an approximate distinction is made between the state spaces of collapsing matter and trapped Hawking partner modes. This is because 
the  Killing field $\frac{\partial}{\partial t}$ corresponding to the black hole symmetry is space-like inside the horizon, and therefore physical states are possible with both positive and negative Killing energies~\cite{horowitz2004black,jacobson2013black}. The collapsing matter could have freely fallen across the horizon from the outside where it has positive killing energy (similar to states in $H_{o}$), and since Killing energy is conserved, the collapsing matter can be identified as states of positive Killing energy, with a corresponding Hilbert space $H_{m}$. This also imply that states with negative Killing energies represent states which can never escape to the exterior, or could never have freely fallen across the horizon, and therefore can only be associated to states localized inside the horizon, i.e., the Hilbert space of the trapped Hawking quanta ($H_{i}$)~\cite{jacobson2013black}. Horowitz and Maldacena also point out that the said symmetry is only approximate for an evaporating black hole, which makes the factorizability of $H_{m}$ with $H_{i}$ only approximate~\cite{horowitz2004black}.

A desired resolution to the quantum information problem can now be addressed from the perspective of an external observer who assumes local quantum field theory is valid, and therefore sees a unitary evolution of quantum states between the Hilbert spaces $H_{m}\rightarrow H_{o}$, described by a scattering matrix $\mathcal{S}$ which is unitary. Note that this is different from standard description of scattering problems -- where the asymptotic incoming and outgoing modes are described as modes in the same Hilbert space -- due to the presence of an event horizon. An incoming mode from the asymptote can freely fall across the horizon and become states in the Hilbert space $H_{m}$ of collapsing matter at the interior of a black hole, which is different from $H_{o}$. Indeed, the final state projection approach arrives at the desired solution where the time evolution $|\psi_{m}\rangle\rightarrow |\psi_{o}\rangle$ is described by a unitary scattering matrix by imposing a final state boundary condition at the black hole singularity, but importantly, Horowitz and Maldacena conclude their paper by noting that the story is only complete when the precise mechanism to describe this evolution is available~\cite{horowitz2004black}. We consider this end-note from Horowitz and Maldacena as an important pretext to the present article.
\begin{figure*}
\includegraphics[scale=0.35]{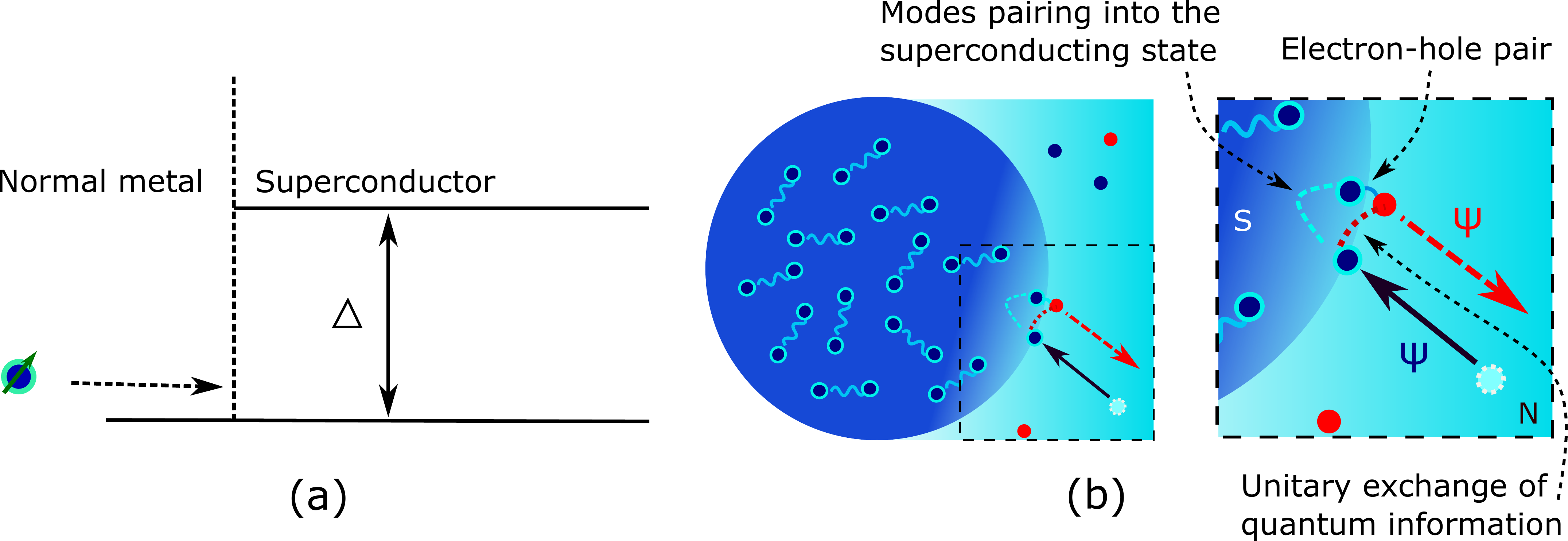}\caption{Andreev reflections from a superconducting condensate. (a) Scattering description of Andreev reflections. Here the superconducting region is described as a potential barrier of amplitude $\Delta$, the superconducting energy gap. (b) Pairing dynamics in Andreev reflections: an electron-like quasiparticle (blue) is retro-reflected as a hole-like quasiparticle (red), while contributing a Cooper pair to the superconducting condensate. In the process, the spin quantum information (encoded in $\psi$) in the incoming electron is transferred to the outgoing hole.  Note that the present article discusses a simplified one dimensional model~\cite{nakano1994second}.
\label{fig1}}
\end{figure*}

Additionally, Horowitz-Maldacena model considers a fixed geometry in which black hole evaporation is a slow process where the quantum fluctuations do not change the energy of the final state. They speculate that the the final state should have an associated entropy of the same order of the black hole entropy measured by an external observer. We note that this ansatz by Horowitz and Maldacena also translates to our Andreev reflection analogy, as discussed in the subsequent sections. 
\section{Mode conversion at black hole horizons: A quantum prescription}
We now re-visit mode conversion processes at the event horizon of a black hole, treated as analogous to Andreev reflections in a normal-metal--superconductor interface (see Fig.~\ref{fig1}). In doing so, we adopt a Hamiltonian description for mode conversions at the event horizon, motivated by the Nakano and Takayanagi approach to describe Andreev reflections~\cite{nakano1994second}, which allows us to incorporate the effect of quantum fluctuation of the black hole final state (superconducting quantum ground state in the analogy) on Andreev reflections from a microscopic quantum physics perspective. It should be noted that a few different approaches have been used to describe Andreev reflections in the past~\cite{kummel1969dynamics,mathews1978quasiparticle,blonder1982transition,nagato1993theory,artemenko1978excess,andreev1964thermal,artemenko1979theory,artemenko1979excess,klapwijk2004proximity,dolcini2009andreev}. The final state projection approach used to describe Andreev reflections in Refs.~\cite{manikandan2017andreev,manikandan2018bosons}, treats the effect of the condensate on Andreev reflections as imposing a final state boundary condition on the infalling modes, motivated by the black hole final state projection models. The Nakano and Takayanagi~\cite{nakano1994second} approach to describe Andreev reflections provides a more detailed microscopic description with some crucial similarities to applying a final state boundary condition, that are highlighted in the subsequent discussions.

Nakano and Takayanagi suggest a microscopic Hamiltonian to describe mode conversions at a normal metal superconductor interface, where the factorization of Hilbert spaces is evident in terms of individual modes. We stick to the one dimensional model for simplicity. The effective Hamiltonian describing the interface in one dimension (considering excitations below the superconducting gap that lead to Andreev reflections) in the Nakano and Takayanagi model is given by~\cite{nakano1994second,guinea1988quantum},
\begin{widetext}
\begin{eqnarray}
\mathcal{H}_{\text{eff}}&=&\mathcal{H}_{b}+\mathcal{H}_{I},~~\text{where}~~\mathcal{H}_{b}=-\frac{\lambda^{2}}{2}\sum_{k,\sigma}B_{k}C_{k,\sigma}^{\dagger}C_{k,\sigma},~~\text{and},\nonumber\\
    \mathcal{H}_{I}&=&\lambda^2\sum_{k>0}A_{k,-k}[P^{\dagger}(C_{k\uparrow}C_{-k\downarrow}+C_{-k\uparrow}C_{k\downarrow})+\text{h.c.}]+\gamma \sum_{k>0}(C_{-k\sigma}^{\dagger}C_{k\sigma}+\text{h.c.}).\label{ham}
\end{eqnarray}
\end{widetext}
  A similar Hamiltonian was also suggested as a phenomenological model to describe quantum fluctuations in a normal metal/superconductor interface by Guinea and Sch\"{o}n in~\cite{guinea1988quantum}. The Hamiltonian $\mathcal{H}_{\text{eff}}$ approximates the interaction Hamiltonian of the interface up to unitary transformations, and making the wave bundle approximation on the normal side~\cite{nakano1994second}. Here operators labeled by $C$ annihilate wave bundle states on the normal side. We denote the Hamiltonian terms describing Andreev reflections and ordinary reflections from the interface, using $\mathcal{H}_{I}$. Although we proceed considering terms in $\mathcal{H}_{I}$ as a phenomenological model to describe Hawking radiation in our Andreev reflection analogy, note that Nakano and Takanayagi also estimates the coefficients in $\mathcal{H}_{\text{eff}}$ in their one dimensional model~\cite{nakano1994second}; The couplings $\lambda$ and $\gamma$ depends on the density of states per unit length of the electrode $N(0)$, length of the electrodes $a$, and the bare transmission ($t$) and reflection ($r$) coefficients of the electrodes in contact, identified via the relations~\cite{nakano1994second},
\begin{equation}
    \lambda=t\sqrt{N(0)a\delta\varepsilon},~~\text{and}~~\gamma=\delta\varepsilon N(0)ra.
\end{equation}
The average energy of a freely propagating wave bundle  on the normal side is  $\varepsilon_{k}$ (relative to the Fermi energy), with a spread $\delta\varepsilon$. The functions $A_{k,-k}$ and $B_{k}$ additionally depends on the superconducting gap $\Delta$ and the energy $\varepsilon_{k}$ via relations~\cite{nakano1994second},
\begin{eqnarray}
    A_{k,-k}&=&\frac{\Delta N(0)}{\sqrt{\Delta^{2}-\varepsilon_{k}^{2}}}\arccos{\bigg[\frac{\Delta-\varepsilon_{k}}{2\Delta}\bigg]}^{\frac{1}{2}},\nonumber\\B_{k}&=&\text{const.}-\frac{N(0)\text{ln}(\Delta-\varepsilon_{k})}{2}.\label{coeff}
\end{eqnarray}
The fluctuations of the condensate in the Hamiltonian is captured by the operator $P^{\dagger}$ approximated as~\cite{nakano1994second},
\begin{equation}
P^{\dagger}=J^{-1}\sum_{k}d_{k\uparrow}^{\dagger}d_{-k\downarrow}^{\dagger}\approx e^{-i\hat{\phi}},\label{eqp}
\end{equation} which creates a Cooper pair in the superconducting condensate. An important difference with standard treatments of Andreev reflections is that the phase is treated as an operator $\hat{\phi}$, conjugate to the charge operator $\hat{\mathcal{Q}}$, i.e.,
\begin{equation}
    [\hat{\mathcal{Q}},\hat{\phi}]=\frac{2e}{i},
\end{equation}
making it evident that the operator $P^{\dagger}\approx e^{-i\hat{\phi}}$ changes the charge across the superconductor--normal metal interface by $2e$ upon Andreev reflection~\cite{guinea1988quantum}. We denote the electronic creation operators on the superconducting side with $d^{\dagger}$, and $J$ is the maximum number of states Cooper paired electrons occupy in the condensate, $J\approx N(0)a\hbar\omega_{D}$. It is also important to note that the pair creation operator $P^{\dagger}$ is associated with an interesting angular momentum algebra of Anderson's psuedospin observables describing the superconducting condensate~\cite{nakano1994second,anderson1959theory}, with associated total angular momentum $J$. Therefore $P^{\dagger}$ corresponds to a macroscopic, many-body quantum operator of the condensate. The Hamiltonian $\mathcal{H}_{I}$ connects this many-body quantum operator of the condensate to quasiparticle modes at the interface. A Cooper pair is always exchanged with the condensate when mode conversions occur, where the pair creation/annihilation operator permits the description of an addition/removal of a single Cooper pair with the condensate, without changing the quasiparticle occupancy of the condensate~\cite{nakano1994second,anderson1959theory}.

This identification is also useful to comment on how the condensate as a whole can be thought to influence mode conversion processes at the boundary, as discussed in the final state projection approach to describe Andreev reflections~\cite{manikandan2017andreev}; Here the Hamiltonian $\mathcal{H}_{I}$ makes it evident that the condensate imposes a certain pairing symmetry for the infalling modes as $P^{\dagger}$ has singlet symmetry, while effectively swapping the quantum information to an outgoing mode via local interactions. The interaction is mediated via the condensate, making Andreev reflections a non-trivial mode swapping with some  similarities to optical phase conjugation in a third order non-linear medium, via four-wave mixing~\cite{beenakker2000does}; here the incoming electron in Andreev reflections can be thought of as the signal beam in four wave mixing, and the outgoing hole is analogous to the retro-reflected conjugate beam, while modes pairing in the condensate mimic the pump beams. In the bosonic case, this analogy is exact. The final state projection model where the condensate is treated as applying a singlet pairing symmetry for the infalling modes circumvents this detailed dynamics, but contains the essential non-trivial back-action of the condensate on Andreev reflections.

 It is worth mentioning that the Hamiltonian $\mathcal{H}_{I}$ is not fully perfect to describe Andreev reflections~\cite{nakano1994second}; for instance, $\mathcal{H}_{I}$ describes Andreev reflections as fully momentum conserving, time-reversal symmetric process, while in reality, Andreev reflections are not fully momentum conserving on the normal side. This is because the momentum of the Cooper pair generated in the Andreev reflection process is not fully determined in $\mathcal{H}_{I}$~\cite{nakano1994second}. In the following, we may use a phenomenological modification to $\mathcal{H}_{I}$ to partially address this issue, where we replace the matrix element of $A$ with $A_{\kappa,q}$ which couple wave vectors $\kappa,~q$. While replacing $q\rightarrow -\kappa$ reduces to the above case~\cite{nakano1994second}, it is known from experiments that Andreev reflections, in reality, corresponds to choices of $\kappa,~q$ such that momentum is only approximately conserved; here  $\kappa$ is the wave vector of an incoming electron-like mode, $\kappa=k_{F}+\delta k$ and $q=-k_{F}+\delta k$, where $k_{F}$ is the Fermi wave vector. Note that in such a phenomenological modification,  $A_{\kappa,q} \rightarrow A_{\kappa,-\kappa}$ in the limit $\delta k\rightarrow 0$, and therefore the Andreev reflection amplitudes $A_{\kappa,q}$ may still be approximated in this limit using Eq.~\eqref{coeff}. A physically relevant scenario where the change in momentum upon Andreev reflections is negligible is when $\Delta\ll E_{F}$~\cite{beenakker2000does}; here Andreev reflections occur as a fully momentum conserving process, where $\mathcal{H}_{I}$ in Eq.~\eqref{ham} tends to be exact.

In particular, we are interested in the time evolution of an arbitrary incoming electron-like mode at the interface, 
\begin{equation}
    \psi_{\kappa,e}^{\dagger}(0)=\alpha C^{\dagger}_{\kappa\uparrow}+\beta C^{\dagger}_{\kappa\downarrow},
\end{equation}
where $\psi^{\dagger}$ encodes quantum information about the amplitudes $\alpha$ and $\beta$ of the spin state,
\begin{equation}
    |\psi\rangle= \alpha\vert\uparrow\rangle+\beta\vert\downarrow\rangle,
\end{equation} in the wave bundle mode denoted by wave vector $\kappa$. In the analogy to black hole context, $\psi^{\dagger}_{\kappa,e}$ describes the quantum information encoded in the incoming mode, which is subsequently freely falling across the horizon, mode-converted as the collapsing matter. The conversion of mode $\psi_{\kappa,e}^{\dagger}$ at the interface is determined by $\mathcal{H}_{I}$, described by the dynamical equation,
\begin{widetext}
\begin{equation}
    i\hbar\frac{d\psi_{\kappa,e}^{\dagger}}{dt}=-\gamma(\alpha C_{-\kappa\uparrow}^{\dagger}+\beta C_{-\kappa\downarrow}^{\dagger})+\lambda^{2}A_{\kappa,q}P^{\dagger}(\alpha C_{q\downarrow}-\beta C_{q\uparrow})=-\gamma \psi_{-\kappa,e}^{\dagger}+\lambda^{2}A_{\kappa,q}P^{\dagger}\psi_{-q,h}^{\dagger}.\label{timEv}
\end{equation}
\end{widetext}
 The presence of an energy gap in the superconductor necessitates that the incoming electron has to be either ordinarily reflected or Andreev reflected.  The Hamiltonian evolution at the interface shown in Eq.~\eqref{timEv} demonstrates this physics, where an infalling mode in an arbitrary quantum spin state is mapped into an ordinarily reflected electron  $\psi_{-\kappa,e}^{\dagger}$ (note that the momentum is changed $\kappa\rightarrow -\kappa$), and an Andreev reflected hole $\psi_{-q,h}^{\dagger}$(defined as a hole-like excitation w.r.t the quasiparticle vacuum, see Appendix A). Note that the spin quantum information is manifestly preserved. Additionally, the operator $P^{\dagger}$ indicates that a Cooper pair has been added to the superconducting  condensate.

One can also consider an ideal interface such that reflectivity is zero. In this case, the quantum information is fully Andreev reflected, 
\begin{equation}
    \psi^{\dagger}_{\kappa,e}\rightarrow\psi^{\dagger}_{-q,h},
\end{equation}
by addition of a Cooper pair into the condensate (see Appendix B). In the analogy to the black hole context, $\psi^{\dagger}_{-q,h}$ describes the quantum information encoded in the outgoing Hawking quantum. 

We now discuss how the first term in Hamiltonian $\mathcal{H}_{I}$ which describes Andreev reflections, naturally captures the state dynamics between the incoming and outgoing modes as a mapping between different Hilbert spaces involved, as Horowitz and Maldacena describe. Here, a Cooper pair is created by mode-converting the incoming electron (positive excitation energy w.r.t to the Fermi level $\rightarrow$ analogous to the incoming particle freely falling across the horizon, mode-converted as the collapsing matter in $H_{m}$), and an infalling, electron-like excitation from the interface (negative excitation energy w.r.t the Fermi level $\rightarrow$ analogous to the trapped Hawking quantum in $H_{i}$). The coupling via the Cooper pair creation operator in $\mathcal{H}_{I}$ implies that the quantum information traverses via the condensate -- analogous to the quantum information encoded in incoming particles freely falling across the horizon as collapsing matter -- before getting Andreev reflected in the outgoing hole (positive excitation energy w.r.t to the hole vacuum $\rightarrow$ analogous to the outgoing Hawking quantum in $H_{o}$). This process where electrons from the normal metal scatter into the condensate via Andreev reflections, causing the superconducting correlations extend slightly into the normal region at the interface, is also known as the superconducting proximity effect~\cite{klapwijk2004proximity}.  The factorizability of Hilbert spaces is also evident in the Andreev reflection paradigm as the infalling and outgoing modes are of different kind of quasi-particle excitations, electron-like and hole-like. 

Additionally, the Andreev reflected hole acquires a phase difference of $-(\pi/2+\phi)$ relative to the incoming electron, where $\phi$ is the macroscopic phase of the condensate (see Appendix B). The description above is also in good agreement with the final state projection approach~\cite{manikandan2017andreev,manikandan2018bosons}, accurately predicting the dynamics of quantum information in Andreev reflections, and the relative phase acquired upon Andreev reflections: the phase $\phi$ of the condensate. Both the predictions are experimentally observable, the spin state of the Andreev reflected electron/hole (using quantum spin state tomography) and the relative phase acquired (possible in an interferometer like setup using an S-N-S junction).

Finally, note that an effective temperature of the Bardeen-Cooper-Schrieffer (BCS) superconducting quantum ground state~\cite{bardeen1957theory} can be derived from entropy considerations, from the spin-partitioned entanglement entropy of the BCS state~\cite{puspus2014entanglement}. For Cooper pairs added to the condensate upon Andreev reflections, the spin partitioned entanglement entropy accounts for the increase in entropy of an incoming electron as it enters the superfluid due to BCS pairing. This entropy increase is similar to that experienced by an infalling observer in the final state proposal, who can only access parts of the black hole interior $H_{m}\otimes H_{in}$~\cite{horowitz2004black}. The entropy spectrum of the BCS state is peaked about the Fermi level, where Andreev reflections dominate.  Also note that Andreev reflections occur as a momentum conserving process in the limit $\Delta\ll E_{F}$, as the change in momentum is negligible~\cite{beenakker2000does}. In this limit, the entropy of the BCS ground state also scales as an area -- the area of the Fermi surface~\cite{puspus2014entanglement}. The associated effective temperature is almost equal to the critical temperature of the superconductor,
\begin{equation}
    T_{c}=\frac{\Delta}{1.76 k_{B}}\approx 1.13\frac{\hbar v_{F}}{4 \pi r k_{B}}.
\end{equation}
Here $\Delta$ is the superconducting gap energy, $\hbar$ is Planck's constant, $k_{B}$ is Boltzmann's constant, and $r=\lambda/2$  where $\lambda$ is the superconducting coherence length, and $v_{F}$ is the Fermi velocity~\footnote{As has already been pointed out~\cite{manikandan2017andreev}, the temperature scales similar to the temperature of a Schwarzschild black hole, where $v_{F}\rightarrow c$, the speed of light, and $r\rightarrow r_{s}$, the Schwarzschild radius.}. The entropy of Andreev reflections can also be calculated by considering electrons/ holes at the superconductor/normal metal interface, within a favourable energy range (around the Fermi energy) to participate in Andreev reflections. The entropy spectrum of electrons/holes at $T_{c}$ is almost identical to that of the BCS ground state discussed before~\cite{puspus2014entanglement}, but can now be understood as the entropy measured by an external observer who is ignorant about the microscopic dynamics and the spin quantum state of Andreev reflecting quasiparticles, and therefore treats Andreev reflection as a thermionic emission process from the interface. 
\subsection{Comparison to alternate treatments of the black hole quantum information paradox}
 We now compare our resolution to the black hole information problem to some of the recent approaches.  A comprehensive discussion of several notable treatments of the black hole information problem comparable to the final state proposal can be found in~\cite{chakraborty2017black}. On the information theoretic frontier, quite a lot of research has been motivated by the initial work of Page~\cite{page1993average,page1993information}, followed by the work of Preskill and Hayden~\cite{hayden2007black}, who pioneered the idea that black holes, rather than destroying quantum information, scramble the quantum information unitarily. This suggested that the quantum information encoded in qubits collapsing into a black hole can be retrieved from the emitted Hawking radiation at later times; in particular, any information entering a black hole past the ``half-way point" -- where half of the black hole's initial entropy has been radiated -- should reveal itself rapidly in the emitted Hawking radiation~\cite{hayden2007black}. Subsequently black holes were also conjectured to be the fastest information scramblers in nature~\cite{sekino2008fast,susskind2011addendum}, and motivated further proposals and interesting experiments related to quantum information scrambling~\cite{landsman2019verified,yoshida2017efficient,blok2020quantum,penington2019entanglement,piroli2020random}.  

Despite the rich many-body physics being discussed, such bottom-up approaches to investigating the black hole information problem are rather simple to comprehend. For instance, a more recent implementation of the Hayden and Preskill protocol was presented in~\cite{piroli2020random}; the proposal describes the black hole as a subsystem of microscopic degrees of freedom modeled using $d-$level quantum systems (qudits). The internal scrambling dynamics is described using a random two-body unitary applied between a randomly chosen pair of internal qudits, with probability $p_{1}$. A randomly chosen internal qudit may also interact with an environmental qudit, via a unitary swap operation. The probability of the swap operation is taken to be $p_{2}\leq 1-p_{1}$, such that at each step, nothing happens with probability $1-p_{1}-p_{2}$. By adding an extra quantum dit of information to the subsystem (the black hole), the authors probe the scrambling time, and the time required to retrieve the information injected to the subsystem, for different degree of knowledge about the initial state of the subsystem. For an observer having access to early Hawking radiation, the authors show that information can be retrieved rather quickly by performing measurements on the environmental qudits, in agreement with the Hayden and Preskill description of a black hole past the half-way point.  

Such a quick retrieval of quantum information in late stages of the evaporation process can be understood as resulting from an entanglement swapping operation~\cite{lloyd2014unitarity}, which also resolves an entanglement monogamy issue for an evaporating black hole -- between the early Hawking radiation, the black hole, and the new Hawking radiation~\cite{almheiri2013black,lloyd2014unitarity}. This aspect of the information problem was revisited recently in~\cite{agarwal2019toy} using a unitary circuit model with additional decoherences implemented as projective quantum measurements; their results indicate a possible resolution of the paradox around the half-way point, but the paradox possibly re-appears at much later stages of the evaporation process where the black hole reduces to a Planck size remnant~\cite{agarwal2019toy}. 

In comparison to the approaches in~\cite{piroli2020random,hayden2007black,blok2020quantum,landsman2019verified,yoshida2017efficient,penington2019entanglement} which also allows to simulate the early stages of an evaporating black hole, our analogy is more suitable for a black hole in its late stages of the evaporation process (the final state)~\cite{piroli2020random,hayden2007black,horowitz2004black}, where the incoming quantum information is revealed rapidly in Andreev reflected quasi-particles. Our model is also devoid of the entanglement monogamy issue~\cite{almheiri2013black,lloyd2014unitarity,agarwal2019toy}, and this follows from the microscopic Hamiltonian $\mathcal{H}_{I}$ given in Eq.~\eqref{ham}, which swaps entanglement unitarily when initial quantum correlations are present. 

Finally, recall that the black hole final state approach~\cite{horowitz2004black} was an attempt to see what simple modifications in Hawking's original calculation for the black hole interior can resolve the information problem, given inputs from various proposals including the AdS/CFT correspondence~\cite{maldacena1999large} which definitively suggested that there cannot be any information loss from the point of view of the boundary conformal field theory (CFT) which is unitary~\cite{penington2019entanglement}. The modification Horowitz and Maldacena suggested was to impose a final state boundary condition at the singularity; a maximally entangled choice for the final state ensures that no information is stuck at the singularity, and subsequently there is no information loss. In comparison, recent alternate approaches to providing a consistent bulk/interior description can be found in~\cite{penington2019entanglement,akers2019simple,almheiri2019entropy,czech2012gravity} using techniques such as entanglement wedge reconstruction~\cite{czech2012gravity,headrick2014causality,wall2014maximin}. The similarity is that their descriptions of the bulk are holographically dual to the Hayden and Preskill protocol~\cite{hayden2007black}, and the latter permits an account within the final state proposal as a simple entanglement swapping~\cite{lloyd2014unitarity}. This is also an important connection to make as the original proposal of Hayden and Preskill describes unitarity from the boundary CFT point of view~\cite{penington2019entanglement}. 

The final state proposal, despite being simple, had several shortcomings which made it less admirable, but in the remainder of this article we discuss how our Andreev reflection analogy resolves some of these important shortcomings of the final state proposal, which also makes our analysis timely and relevant.

\section{Implications to the black hole final state proposal}

We now discuss the implications of our results for the black hole final state proposal~\cite{horowitz2004black,lloyd2014unitarity,gottesman2004comment}. First, note that the present description reproduces major conclusions of the final state projection approach without having to discuss various quantum correlations across or within the condensate. The condensate is indeed described as a superfluid of pairs, but the effect of the superfluid state on Andreev reflections is treated differently in the present approach. We used an effective Hamiltonian and time evolution of incoming modes in Eq.~\eqref{timEv} to arrive at the transfer of quantum information in the mode conversion process. Note that this also allows us to comment on the speed at which information traverses the condensate. The microscopic description presented in this article precisely corresponds to the local interactions that mediate information transfer in Andreev reflections, as discussed in~\cite{manikandan2018bosons,manikandan2017andreev}. Therefore the maximum speed at which the information traverses is roughly limited by the speed of sound in the lattice, as the superconducting pairing interactions are mediated by lattice phonons.

It was conjectured in~\cite{manikandan2017andreev,manikandan2018bosons} that the macroscopic quantum final state of black hole in the final state projection models~\cite{lloyd2014unitarity,horowitz2004black} can be treated as the superfluid quantum ground state of fermions and bosons respectively. For spin-half fermions, this has the form of the Bardeen-Cooper-Schrieffer (BCS) state~\cite{bardeen1957theory},
\begin{equation}
    |\Psi\rangle = \prod_{k}(u_{k}+v_{k}e^{i\phi}d^{\dagger}_{k\uparrow}d^{\dagger}_{-k\downarrow})|0\rangle.
\end{equation}
The coefficients $u_{k}$ and $v_{k}$ are determined in BCS theory via relations,
\begin{equation}
    u_{k}=\cos\frac{\theta_{k}}{2}~~\text{and}~~v_{k}=\sin\frac{\theta_{k}}{2},
\end{equation}
where $\sin\theta_{k}=\frac{\Delta}{\sqrt{\Delta^{2}+\varepsilon_{k}^{2}}}$, and $\cos\theta_{k}=\frac{\varepsilon_{k}}{\sqrt{\Delta^{2}+\varepsilon_{k}^{2}}}$, for the superconducting gap energy $\Delta$. The analogy was primarily built on information considerations, based on how Andreev reflections preserve the quantum information by transferring them to the outgoing modes. An additional well known quality the BCS wavefunction possess is its off-diagonal long range ordering, which gives a sense of rigidity to the macroscopic quantum final state~\cite{rensink1967off,manikandan2017andreev}. The present microscopic treatment allows us to take a step forward from the final state proposal, and associate a microscopic Hamiltonian presented in Eq.~\eqref{ham} to describe the dynamics of mode conversion processes at the event horizon. The Hamiltonian $\mathcal{H}_{I}$ has a remarkable feature that it connects the wave bundle operators $C_{k}$ on the normal side to a macroscopic, many-body quantum operator $P^{\dagger}$ of the condensate that describe exchange of quasiparticles with the final state.

We emphasize, based on the microscopic analysis presented in this article, that the analogy to Andreev reflections appears to resolve two of the noted problems in the quantum description of a black hole, (1) the Trans-Planckian reservoir problem already discussed by Jacobson~\cite{jacobson1996origin}, and (2) the black hole information problem. Therefore, we add that the Hamiltonian in Eq.~\eqref{ham} that describes quantum fluctuations in a superconductor--normal metal interface~\cite{nakano1994second,guinea1988quantum}, has the desired properties to describe the mode conversion processes occurring at the event horizon of a black hole causing the black hole to evaporate unitarily. From the microscopic perspective, the evaporation process can be understood as caused by Andreev reflections of incoming hole-like quasiparticles (described by the term in $\mathcal{H}_{I}$ proprortional to $P$) where Cooper pairs are effectively removed from the condensate. 

We now proceed to discuss how the Andreev reflection analogy helps to address major concerns regarding the computational enhancements in the final state approach~\cite{almheiri2013black,bao2016grover}.  Subsequently, we revisit a possible scenario of apparent loss of quantum information in our Andreev reflection analogy, resulting from the quantum information traversing across the condensate via crossed Andreev reflections~\cite{deutscher2002crossed,manikandan2017andreev}.

\subsection{Comment on the computational advantages of the final state proposal}
  Some of the computational enhancements the final state proposal suggest~\cite{lloyd2014unitarity,almheiri2013black,PhysRevLett.106.040403,PhysRevD.84.025007,bao2016grover,brun2012perfect} are a cause of concern, as they could possibly be unphysical ~\cite{almheiri2013black,bao2016grover}; for instance, it has been shown that small departures from unitarity can allow superluminal signalling, and also may allow computations of NP hard problems (class of problems where a given solution can be verified in polynomial time, while exponentially many solutions are possible) in polynomial time. This  would suggest a boost over the query complexity lower bound for search algorithms~\cite{bennett1997strengths} saturated by the Grover's search protocol~\cite{grover1996fast,bao2016grover}.  Additionally, it has been pointed out that the final state projection at the black hole singularity may allow backward in time signalling even if the scattering matrix is unitary~\cite{almheiri2013black,lloyd2014unitarity}; suggested resolutions to this puzzle relies on the complexity of performing such a computational task, which requires fast and efficient decoding of information from the outgoing Hawking radiation~\cite{lloyd2014unitarity,harlow2013quantum}. Some aspects of the origin of these computational advantages in the Horowitz-Maldacena model can be attributed to the ability to post-select on certain outcomes in a quantum measurement process, which can allow a quantum computer to solve a class of problems in polynomial time~\cite{aaronson2005quantum}; this aspect has also been discussed in the context of some variations of the Horowitz-Maldacena model, proposed to describe closed time-like curves in quantum mechanics~\cite{PhysRevLett.106.040403,PhysRevD.84.025007,brun2012perfect}. 

While the enticing similarities to the final state proposal are crucial to the analogy we discussed, the important progress we make -- also relevant to the quantum computational aspects of the final state proposal -- is that Andreev reflections suggests a possible microscopic treatment from which the final state proposal can emerge, devoid of the computational enhancements resulting from a final state projection. In our microscopic approach, the unitarity of evolution follows from a microscopic interaction Hamiltonian at the event horizon of the type $\mathcal{H}_{I}$, and Eq.~\eqref{timEv}, unlike the final state projection approach where unitarity can only be discussed at the level of a scattering matrix~\cite{horowitz2004black,lloyd2014unitarity}. As a consequence, in the Andreev reflection setting, we do not expect any computational enhancements beyond the limits set by a standard quantum computer with linear and unitary operations.

 In spite of that, our analogy suggests some practical advantages with possible quantum device applications, inspired from the black hole final state proposal. Superconductors used as ``quantum information mirrors" can indeed enhance quantum information processing and quantum computing tasks within the limitations of a standard quantum computer, in the presence of additional decoherences; clever device architectures can be constructed using sandwiches of normal metallic electrodes and superconductors, which can significantly increase the life time of spin qubits using multiple Andreev reflections. This could lead to remarkable advances for matter-based spin qubit platforms, where short lifetime of a qubit~\cite{dial2013charge,watson2017atomically,baart2016single,tyryshkin2012electron,hanson2007spins,scarlino2014spin} is a critical problem to be addressed.
 
\begin{figure}
\includegraphics[scale=0.3]{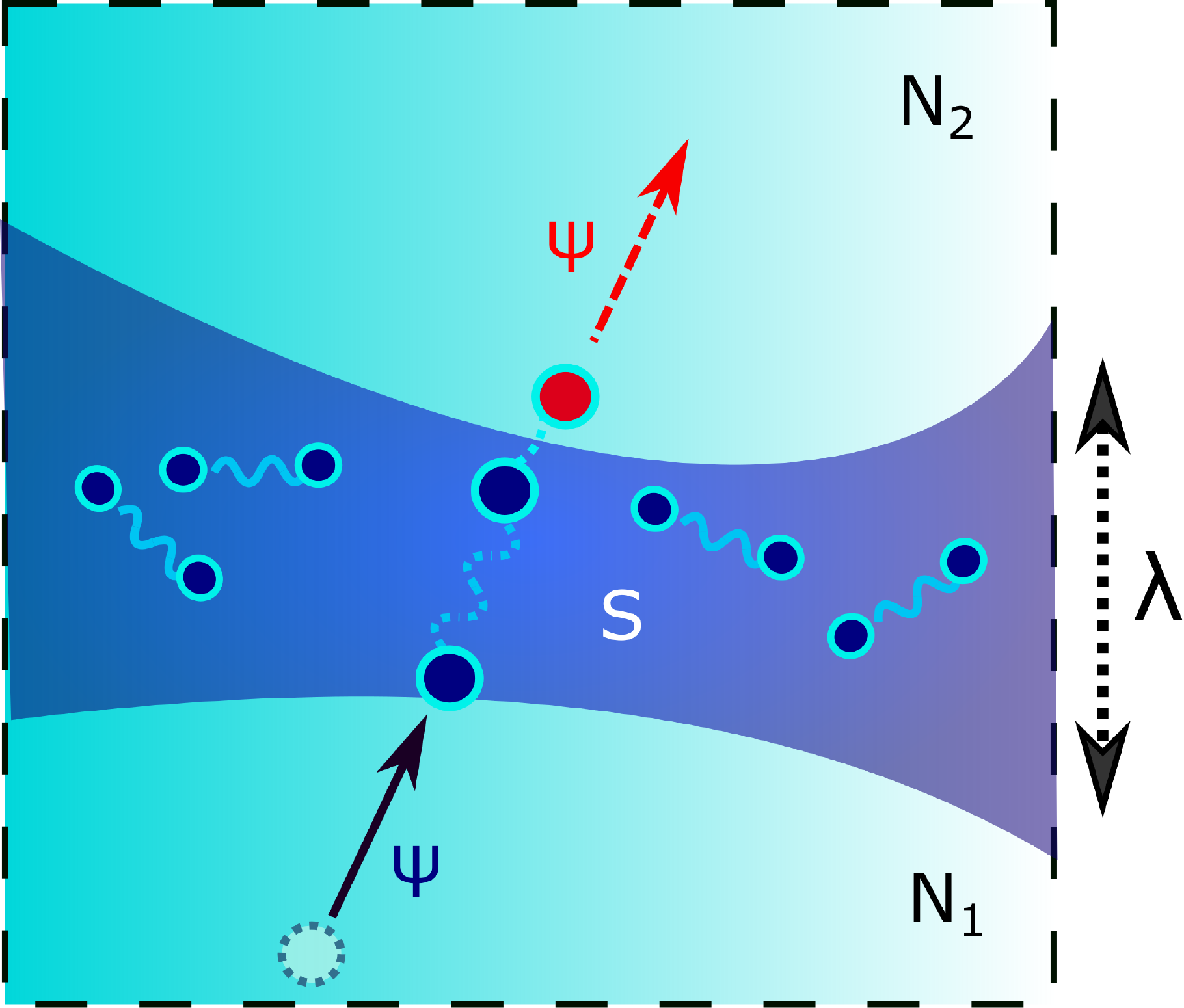}\caption{Crossed Andreev reflections: When a superconductor ($S$) having width comparable to the superconducting coherence length $\lambda$ is sandwiched between two normal metals ($N_{1}$ and $N_{2}$), Andreev reflection can happen across a superconducting condensate, where an incoming electron in $N_{1}$ is Andreev reflected as a hole in $N_{2}$. The spin quantum information (encoded in $\psi$) which is apparently lost in $N_{1}$ reappears in $N_{2}$~\cite{manikandan2017andreev}.\label{fig2}}
\end{figure}

\subsection{Apparent loss of quantum information}

Before we conclude, we discuss a possible generalization of our Andreev reflection analogy which features an apparent loss of quantum information locally, although the quantum information is globally preserved. To this effect, we consider a superconductor of width comparable to the coherence length of a superconductor, $\lambda$, sandwiched between two normal metals. Assuming both interfaces to be ideal (reflectively zero), we can write the following Hamiltonian for incoming modes having energies below the superconducting gap $\Delta$, 
\begin{eqnarray}
    \mathcal{H}_{I}^{ij} &=& [A^{ij}_{\kappa,q}(C_{\kappa\uparrow}^{N_{i}}P^{\dagger}C_{q\downarrow}^{N_{j}}+C_{q\uparrow}^{N_{j}}P^{\dagger}C_{\kappa\downarrow}^{N_{i}})+\text{h.c.}].
\end{eqnarray}

Here combinations of $i,j=1,2$ represent the different combinations of possible Andreev reflections possible, all preserving the spin quantum information $\psi^{\dagger}_{\kappa,e}\leftrightarrow\psi^{\dagger}_{-q,h}$; direct Andreev reflections are described by $i=j$ where Andreev reflected mode is produced on the same side as the incoming mode, while $i\neq j$ describes Andreev reflections across the superconductor (crossed Andreev reflections~\cite{deutscher2002crossed}) where spin quantum information is apparently lost on one side of the superconductor, but re-appears on the quasiparticle mode Andreev reflected on the other side. See Fig.~\ref{fig2}. 

As a naive extension of our analogy, we note that such a process describes information transfer possible in an Einstein-Rosen bridge (a wormhole)~\cite{einstein1935particle}, where the quantum information is apparently lost in one horizon as a result of traversing the wormhole~\cite{manikandan2017andreev}. A similar construction, where the microstates of an Einstein-Rosen bridge is built by pairing microstates of two black holes, has also been discussed by Maldacena and Susskind in~\cite{maldacena2013cool}.  
 
\section{Conclusions}
We presented a microscopic description of Hawking radiation as Andreev reflections where the unitarity of information transfer is evident, without having to rely on the assumptions of the final state projection approach. We find good agreement with the predictions of the final state projection approach, substantiating that the latter is indeed a good description when the physics of Hawking radiation is treated as a scattering problem, where the microscopic details of the scattering center are irrelevant. Nevertheless, we note that the alternate description fully relying on microscopic dynamical laws describing the superfluid state allowed us to resolve important shortcomings of the final state projection approach, pertaining to departure from unitarity, and possibly unphysical quantum computational enhancements resulting from final state projection~\cite{almheiri2013black,bao2016grover,lloyd2014unitarity,gottesman2004comment,brun2012perfect,PhysRevD.84.025007,PhysRevLett.106.040403,aaronson2005quantum}.

 Therefore, the present analysis further strengthens the conjecture that the black hole final state could be a superfluid. Naively, it is tempting to associate a simple many body quantum final state to a black hole as we know that classically, its mass, charge and angular momentum completely describe a black hole. Superfluid condensates have additional benefits; Apart from the fact that they are described by very few parameters such as the average particle density and a macroscopic phase, the necessary microscopic details provided here re-enforce our previous proposal that they also resolve the famous black hole information paradox by acting like a mirror to quantum information. This, together with an earlier observation by Jacobson that Andreev reflections can also resolve the Trans-Planckian problem at the event horizon~\cite{jacobson1996origin}, makes it a strong candidate description of the quantum physics at black hole horizons.  

 Yet another important progress we make is that we conjecture a Hamiltonian $\mathcal{H}_{I}$ presented in Eq.~\eqref{ham} to describe mode conversion processes at the event horizon, treated as Andreev reflections. The Hamitonian  $\mathcal{H}_{I}$ has terms describing interactions between a macroscopic many-body quantum operator of the condensate and microscopic quasiparticle modes at the interface, describing how the final state projection model can emerge as an effective description from interactions between the infalling modes and the macroscopic condensate. We also generalized the Hamiltonian $\mathcal{H}_{I}$ to describe mode conversion processes involving an Einstein-Rosen bridge (wormhole), that allows to provide a unitary description of apparent loss of information in an Einstein-Rosen bridge as a result of traversing the wormhole via crossed Andreev reflections.
 
Finally, we address how small deviations in our model may affect the quantum information dynamics in Andreev reflections. First, note that the approach developed by Nakano and Takayanagi to describe Andreev reflections accounts for small fluctuations in energy/momentum by considering wave bundles instead of traveling waves. Therefore the model is immune to small fluctuations in energy/momentum of incoming modes, especially in the limit $\Delta\ll E_{F}$, where the change in momentum upon Andreev reflections is negligible such that $\mathcal{H}_{I}$ in Eq.~\eqref{ham} tends to be exact. Secondly, an incoming mode could be in a superposition of different wave bundle states in the momentum space, labeled by $\kappa$, and in this case our model predicts that the outgoing hole may be in a superposition of different outgoing wave bundle states, determined by coefficients $A_{\kappa,q}$, but still preserving unitarity. Since the momentum of the Cooper pair generated is indeterminate in $\mathcal{H}_{I}$, it ensures that the condensate does not retain any information about the infalling mode, and that the quantum information is fully Andreev reflected. 

 We emphasize that, albeit the similarities we discussed, our analysis does not qualify as an exact correspondence between the two fields; there are obvious differences between a superconductor and a black hole. Nevertheless, the analogy we developed  points at an exciting opportunity that certain quantum theories of gravity can be experimentally tested using superconductor/normal metal interfaces. Conversely, superconductors, used as ``quantum information mirrors" are also promising paradigms for quantum information processing and quantum computing tasks. 
 \section{Acknowledgements}
 We thank Juan Maldacena, Karthik Rajeev, S. G. Rajeev, and Fernando Sols for helpful discussions and insightful comments. This work was supported by the National Science Foundation under Grant No. DMR-1809343.

 \appendix
\section{Quantum spin state of the Andreev reflected hole}
Here we use particle-hole symmetry arguments to determine the quantum spin state of Andreev reflected hole. We define the filled Fermi sea as, \begin{equation}
    |G\rangle =\prod_{|k|<k_{F}}C^{\dagger}_{k\downarrow}C^{\dagger}_{k\uparrow}|0\rangle= |1_{q\downarrow}1_{q\uparrow}...\rangle,
\end{equation}
and implement the following transformation into the hole-picture,
\begin{widetext}
\begin{equation}
(\alpha C_{q\downarrow}-\beta C_{q\uparrow})|G\rangle=\alpha|0_{q\downarrow}1_{q\uparrow}...\rangle+\beta|1_{q\downarrow}0_{q\uparrow}...\rangle =(\alpha h_{-q\uparrow}^{\dagger}  +\beta h^{\dagger}_{-q\downarrow})|0'\rangle=\psi_{-q,h}^{\dagger}|0'\rangle.\label{hol}
\end{equation}
\end{widetext}
We have defined the hole creation operators via the relations,
\begin{equation}
    |0_{q\downarrow}1_{q\uparrow}...\rangle=h_{-q\uparrow}^{\dagger}|0'\rangle~\text{and}~|1_{q\downarrow}0_{q\uparrow}...\rangle= h^{\dagger}_{-q\downarrow}|0'\rangle,
\end{equation}
where $|0'\rangle$ denotes the quasiparticle vacuum in the hole picture. Note that, therefore, $\psi_{-q,h}^{\dagger}$ creates an outgoing hole-like quasiparticle encoding quantum information about the amplitudes $\alpha$ and $\beta$ of the spin state $|\psi\rangle = \alpha|\uparrow\rangle+\beta|\downarrow\rangle$.
 \section{Ideal interface}
We now consider the case of an ideal interface where the coefficient of ordinary reflection is assumed to be zero (note that other scattering coefficients would change as appropriate to preserve unitarity). Andreev reflections in this case can be described by combining the time evolution of modes $\psi^{\dagger}_{\kappa,e}$ and $\psi^{\dagger}_{-q,h}$:
      \begin{equation}
    i\hbar\frac{d\psi_{\kappa,e}^{\dagger}}{dt}=\Omega^{\dagger}\psi_{-q,h}^{\dagger},
    \end{equation}
    and
          \begin{equation}
    i\hbar\frac{d\psi_{-q,h}^{\dagger}}{dt}=\Omega\psi_{\kappa,e}^{\dagger}.
    \end{equation}
We have defined $\Omega^{\dagger}=\lambda^{2}A_{\kappa,q}P^{\dagger}=\lambda^{2}A_{k,q}e^{-i\hat{\phi}}$, where we have approximated the operator $P^{\dagger}$ as $P^{\dagger}\approx e^{-i\hat{\phi}}$~\cite{nakano1994second}.  

In order to make comparison with the standard treatments~\cite{kummel1969dynamics,mathews1978quasiparticle,blonder1982transition,nagato1993theory,artemenko1978excess,andreev1964thermal,artemenko1979theory,artemenko1979excess,klapwijk2004proximity,dolcini2009andreev}, we also make the assumption that the condensate has definite phase, and therefore $\hat{\phi}$ is replaced by $\phi$. We therefore obtain the following second order differential equation,
\begin{equation}
    \frac{d^{2}\psi_{\kappa,e}^{\dagger}(t)}{dt^{2}}=-\frac{1}{\hbar^{2}}\omega^{2} \psi_{\kappa,e}^{\dagger}(t),\label{b1}\end{equation}
where we have defined $\omega=\lambda^{2}A_{\kappa,q}$. Additionally,
          \begin{equation}
    i\hbar\frac{d\psi_{\kappa,e}^{\dagger}}{dt}=\omega e^{-i\phi}\psi_{-q,h}^{\dagger},
    \end{equation}
     determines the time derivative at $t=0$.
     The equation~\eqref{b1} can now be solved, which gives the following solution, 
\begin{equation}
    \psi_{\kappa,e}^{\dagger}(t)=\cos\bigg(\frac{\omega t}{\hbar}\bigg)\psi_{\kappa,e}^{\dagger}(0)-ie^{-i\phi}\sin\bigg(\frac{\omega t}{\hbar}\bigg)\psi_{-q,h}^{\dagger}(0). \label{tim}
\end{equation}
Note that Eq.~\eqref{tim} correctly describes two relative phases picked upon Andreev  reflection, the phase change $-\pi/2$ and the additional phase difference between electrons and holes, which is the macroscopic phase $\phi$ of the condensate.

The quantum spin dynamics in Andreev reflections can also be summarized in terms of an effective Hamiltonian for the interface which maps,\begin{equation}
    H_{\omega}|\psi_{\kappa,e}\rangle=\omega P^{\dagger}|\psi_{-q,h}\rangle,~~H_{\omega}|\psi_{-q,h}\rangle=\omega P|\psi_{\kappa,e}\rangle.\label{ham2}
\end{equation}
We retain the Cooper pair creation operator $P^{\dagger}\approx e^{-i\hat{\phi}}$ in the expressions to emphasize that it is not necessary to assume the phase operator $\hat{\phi}$ takes definite value. We suppress the superconducting state space for simplicity; It is implied that $P^{\dagger}$ is an operator acting on the state space of the superconductor (creating a Cooper pair), where $P^{\dagger}P\approx PP^{\dagger}\approx\mathbb{I}$. The states $|\psi_{\kappa,e}\rangle,~|\psi_{-q,h}\rangle$ can be treated orthogonal on the normal side, as they represent electron-like and hole-like excitation encoded in different modes. In their basis, we can represent $H_{\omega}$ as,
\begin{equation}
    H_{\omega}= \omega\begin{bmatrix}0&&P\\P^{\dagger}&&0   
    \end{bmatrix},~\text{where}~H_{\omega}^{2}\approx\omega^{2}\begin{bmatrix}\mathbb{I}&&0\\0&&\mathbb{I}   
    \end{bmatrix}.
\end{equation}
With this, we find the time evolution of the initial state $|\psi_{\kappa,e}\rangle$ is,
\begin{widetext}
\begin{eqnarray}
    e^{-\frac{i}{\hbar}H_{\omega}t}|\psi_{\kappa,e}\rangle &=& \bigg(1-\frac{i}{\hbar}H_{\omega}t-\frac{H_{\omega}^{2}t^{2}}{2!\hbar^{2}}+i\frac{H_{\omega}^{3}t^{3}}{3!\hbar^{3}}...\bigg)|\psi_{\kappa,e}\rangle\approx|\psi_{\kappa,e}\rangle-\frac{i\omega t P^{\dagger}}{\hbar} |\psi_{-q,h}\rangle-\frac{\omega^{2}t^{2}}{2!\hbar^{2}}|\psi_{\kappa,e}\rangle+\frac{i\omega^{3}t^{3}P^{\dagger}}{3!\hbar^{3}}|\psi_{-q,h}\rangle+...\nonumber\\&=&\bigg(1-\frac{\omega^{2}t^{2}}{2!\hbar^{2}}+...\bigg)|\psi_{\kappa,e}\rangle-iP^{\dagger}\bigg(\frac{\omega t}{\hbar}-\frac{\omega^{3} t^{3}}{3!\hbar^{3}}+...\bigg)|\psi_{-q,h}\rangle\nonumber\\
    &=&\cos\bigg(\frac{\omega t}{\hbar}\bigg)|\psi_{\kappa,e}\rangle-iP^{\dagger}\sin\bigg(\frac{\omega t}{\hbar}\bigg)|\psi_{-q,h}\rangle\approx\cos\bigg(\frac{\omega t}{\hbar}\bigg)|\psi_{\kappa,e}\rangle-ie^{-i\hat{\phi}}\sin\bigg(\frac{\omega t}{\hbar}\bigg)|\psi_{-q,h}\rangle,\label{tim2}
\end{eqnarray}
\end{widetext}
similar to Eq.~\eqref{tim}. Note that for an interaction time $t=\tau\sim \frac{\pi\hbar}{2\omega}$ -- which also satisfies the energy time uncertainty principle for the interface, $\omega\tau>\frac{\hbar}{2}$ -- we have the incoming electron-like mode fully converted into the outgoing hole-like mode, while adding a Cooper pair into the condensate. The hole propagates in the normal region, encoding the spin quantum information $|\psi\rangle$ of the incoming electron. 

The final state projection approach presented in~\cite{manikandan2017andreev} also predicts that the quantum spin state of the outgoing hole is  $|\psi\rangle$. The phase that gets accumulated in the final state approach include a phase factor of sign$(j)e^{-i\frac{\pi}{2}}$ from the tunneling of incoming electron into the condensate, treated as a resonant interaction~\cite{manikandan2017andreev},
\begin{equation}
    H_{C}^{'}|\psi_{e}\rangle=j|\psi_{d}\rangle,~~\text{and}~~H_{C}^{'}|\psi_{d}\rangle=j|\psi_{e}\rangle.
\end{equation}
The time evolved state becomes,
\begin{equation}
    e^{-\frac{iH_{C}^{'}\tau}{\hbar}}|\psi_{e}\rangle = -i\sin\bigg(\frac{j\tau}{\hbar}\bigg)|\psi_{d}\rangle+\cos\bigg(\frac{j\tau}{\hbar}\bigg)|\psi_{e}\rangle.
\end{equation}
The relative phase $\frac{\pi}{2}$ was missed out in~\cite{manikandan2017andreev}. Assuming availability of a singlet electron--hole pair, the projection onto the BCS state for infalling modes adds a phase factor of $-e^{-i\phi}$, where $\phi$ is the phase of the condensate~\cite{manikandan2017andreev}. By choosing sign$(j)=-$sign$(\omega)$, we find that total phase changes that occur in final state approach presented in~\cite{manikandan2017andreev} is equal to $-(\frac{\pi}{2}+\phi)$.
\bibliography{Reference.bib}

\begin{thebibliography}{10}

\bibitem{andreev1964thermal}
A.~F. Andreev, ``Thermal conductivity of the intermediate state of
  superconductors,'' {\em Sov. Phys. JETP}, vol.~19, no.~5, p.~1228, 1964.

\bibitem{kummel1969dynamics}
R.~K{\"u}mmel, ``Dynamics of current flow through the phase-boundary between a
  normal and a superconducting region,'' {\em Zeitschrift f{\"u}r Physik A
  Hadrons and nuclei}, vol.~218, no.~5, pp.~472--494, 1969.

\bibitem{mathews1978quasiparticle}
W.~N. Mathews~Jr, ``Quasiparticle, charge, and energy conservation in
  weak-coupling superconductors,'' {\em physica status solidi (b)}, vol.~90,
  no.~1, pp.~327--338, 1978.

\bibitem{blonder1982transition}
G.~E. Blonder, M.~Tinkham, and T.~M. Klapwijk, ``Transition from metallic to
  tunneling regimes in superconducting microconstrictions: Excess current,
  charge imbalance, and supercurrent conversion,'' {\em Physical Review B},
  vol.~25, no.~7, p.~4515, 1982.

\bibitem{nagato1993theory}
Y.~Nagato, K.~Nagai, and J.~Hara, ``Theory of the andreev reflection and the
  density of states in proximity contact normal-superconducting infinite
  double-layer,'' {\em Journal of low temperature physics}, vol.~93, no.~1-2,
  pp.~33--56, 1993.

\bibitem{artemenko1978excess}
S.~N. Artemenko, A.~F. Volkov, and A.~V. Zaitsev, ``The excess current in
  superconducting point junctions,'' {\em JETP Letters}, vol.~28, pp.~589--591,
  1978.

\bibitem{artemenko1979theory}
S.~N. Artemenko, A.~F. Volkov, and A.~V. Zaitsev, ``Theory of the nonstationary
  josephson effect in short superconducting contacts,'' {\em Zhurnal
  Eksperimental'noi i Teoreticheskoi Fiziki}, vol.~76, pp.~1816--1833, 1979.

\bibitem{artemenko1979excess}
S.~N. Artemenko, A.~F. Volkov, and A.~V. Zaitsev, ``On the excess current in
  microbridges scs and scn,'' {\em Solid State Communications}, vol.~30,
  no.~12, pp.~771--773, 1979.

\bibitem{klapwijk2004proximity}
T.~M. Klapwijk, ``Proximity effect from an andreev perspective,'' {\em Journal
  of superconductivity}, vol.~17, no.~5, pp.~593--611, 2004.

\bibitem{dolcini2009andreev}
F.~Dolcini, ``Andreev reflection,'' {\em Lecture Notes for XXIII Physics
  GradDays}, vol.~5, p.~9, 2009.

\bibitem{beenakker2000does}
C.~W.~J. Beenakker, ``Why does a metal-superconductor junction have a
  resistance?,'' in {\em Quantum mesoscopic phenomena and mesoscopic devices in
  microelectronics}, pp.~51--60, Springer, 2000.

\bibitem{courtois1999spectral}
H.~Courtois, P.~Charlat, P.~Gandit, D.~Mailly, and B.~Pannetier, ``The spectral
  conductance of a proximity superconductor and the reentrance effect,'' {\em
  Journal of low temperature physics}, vol.~116, no.~3-4, pp.~187--213, 1999.

\bibitem{zapata2009andreev}
I.~Zapata and F.~Sols, ``Andreev reflection in bosonic condensates,'' {\em
  Physical review letters}, vol.~102, no.~18, p.~180405, 2009.

\bibitem{beenakker2008colloquium}
C.~W.~J. Beenakker, ``Colloquium: Andreev reflection and klein tunneling in
  graphene,'' {\em Reviews of Modern Physics}, vol.~80, no.~4, p.~1337, 2008.

\bibitem{beenakker2008correspondence}
C.~W.~J. Beenakker, A.~R. Akhmerov, P.~Recher, and J.~Tworzyd{\l}o,
  ``Correspondence between andreev reflection and klein tunneling in bipolar
  graphene,'' {\em Physical Review B}, vol.~77, no.~7, p.~075409, 2008.

\bibitem{klein1929reflexion}
O.~Klein, ``Die reflexion von elektronen an einem potentialsprung nach der
  relativistischen dynamik von dirac,'' {\em Zeitschrift f{\"u}r Physik},
  vol.~53, no.~3-4, pp.~157--165, 1929.

\bibitem{lee2019perfect}
S.~Lee, V.~Stanev, X.~Zhang, D.~Stasak, J.~Flowers, J.~S. Higgins, S.~Dai,
  T.~Blum, X.~Pan, V.~M. Yakovenko, {\em et~al.}, ``Perfect andreev reflection
  due to the klein paradox in a topological superconducting state,'' {\em
  Nature}, vol.~570, no.~7761, pp.~344--348, 2019.

\bibitem{calogeracos1999history}
A.~Calogeracos and N.~Dombey, ``History and physics of the klein paradox,''
  {\em Contemporary physics}, vol.~40, no.~5, pp.~313--321, 1999.

\bibitem{dombey1999seventy}
N.~Dombey and A.~Calogeracos, ``Seventy years of the klein paradox,'' {\em
  Physics Reports}, vol.~315, no.~1-3, pp.~41--58, 1999.

\bibitem{jacobson1996origin}
T.~Jacobson, ``On the origin of the outgoing black hole modes,'' {\em Physical
  Review D}, vol.~53, no.~12, p.~7082, 1996.

\bibitem{manikandan2017andreev}
S.~K. Manikandan and A.~N. Jordan, ``Andreev reflections and the quantum
  physics of black holes,'' {\em Physical Review D}, vol.~96, no.~12,
  p.~124011, 2017.

\bibitem{manikandan2018bosons}
S.~K. Manikandan and A.~N. Jordan, ``Bosons falling into a black hole: A
  superfluid analogue,'' {\em Physical Review D}, vol.~98, no.~12, p.~124043,
  2018.

\bibitem{hawking1975particle}
S.~W. Hawking, ``Particle creation by black holes,'' {\em Communications in
  mathematical physics}, vol.~43, no.~3, pp.~199--220, 1975.

\bibitem{parker2009quantum}
L.~E. Parker and D.~J. Toms, {\em Quantum field theory in curved spacetime:
  quantized fields and gravity}.
\newblock Cambridge university press, Cambridge, UK, 2009.

\bibitem{unruh1981experimental}
W.~G. Unruh, ``Experimental black-hole evaporation?,'' {\em Physical Review
  Letters}, vol.~46, no.~21, p.~1351, 1981.

\bibitem{unruh1995sonic}
W.~G. Unruh, ``Sonic analogue of black holes and the effects of high
  frequencies on black hole evaporation,'' {\em Physical Review D}, vol.~51,
  no.~6, p.~2827, 1995.

\bibitem{hawking1976breakdown}
S.~W. Hawking, ``Breakdown of predictability in gravitational collapse,'' {\em
  Physical Review D}, vol.~14, no.~10, p.~2460, 1976.

\bibitem{horowitz2004black}
G.~T. Horowitz and J.~Maldacena, ``The black hole final state,'' {\em Journal
  of High Energy Physics}, vol.~2004, no.~02, p.~008, 2004.

\bibitem{hayden2007black}
P.~Hayden and J.~Preskill, ``Black holes as mirrors: quantum information in
  random subsystems,'' {\em Journal of High Energy Physics}, vol.~2007, no.~09,
  p.~120, 2007.

\bibitem{lloyd2014unitarity}
S.~Lloyd and J.~Preskill, ``Unitarity of black hole evaporation in final-state
  projection models,'' {\em Journal of High Energy Physics}, vol.~2014, no.~8,
  p.~126, 2014.

\bibitem{gottesman2004comment}
D.~Gottesman and J.~Preskill, ``Comment on``the black hole final state'',''
  {\em Journal of High Energy Physics}, vol.~2004, no.~03, p.~026, 2004.

\bibitem{almheiri2013black}
A.~Almheiri, D.~Marolf, J.~Polchinski, and J.~Sully, ``Black holes:
  complementarity or firewalls?,'' {\em Journal of High Energy Physics},
  vol.~2013, no.~2, p.~62, 2013.

\bibitem{bao2016grover}
N.~Bao, A.~Bouland, and S.~P. Jordan, ``Grover search and the no-signaling
  principle,'' {\em Physical review letters}, vol.~117, no.~12, p.~120501,
  2016.

\bibitem{brun2012perfect}
T.~A. Brun and M.~M. Wilde, ``Perfect state distinguishability and
  computational speedups with postselected closed timelike curves,'' {\em
  Foundations of Physics}, vol.~42, no.~3, pp.~341--361, 2012.

\bibitem{PhysRevD.84.025007}
S.~Lloyd, L.~Maccone, R.~Garcia-Patron, V.~Giovannetti, and Y.~Shikano,
  ``Quantum mechanics of time travel through post-selected teleportation,''
  {\em Phys. Rev. D}, vol.~84, p.~025007, Jul 2011.

\bibitem{PhysRevLett.106.040403}
S.~Lloyd, L.~Maccone, R.~Garcia-Patron, V.~Giovannetti, Y.~Shikano,
  S.~Pirandola, L.~A. Rozema, A.~Darabi, Y.~Soudagar, L.~K. Shalm, and A.~M.
  Steinberg, ``Closed timelike curves via postselection: Theory and
  experimental test of consistency,'' {\em Phys. Rev. Lett.}, vol.~106,
  p.~040403, Jan 2011.

\bibitem{aaronson2005quantum}
S.~Aaronson, ``Quantum computing, postselection, and probabilistic
  polynomial-time,'' {\em Proceedings of the Royal Society A: Mathematical,
  Physical and Engineering Sciences}, vol.~461, no.~2063, pp.~3473--3482, 2005.

\bibitem{nakano1994second}
H.~Nakano and H.~Takayanagi, ``Second-quantization description of andreev
  reflection and the relation to quasiparticle wave approaches,'' {\em Physical
  Review B}, vol.~50, no.~5, p.~3139, 1994.

\bibitem{einstein1935particle}
A.~Einstein and N.~Rosen, ``The particle problem in the general theory of
  relativity,'' {\em Physical Review}, vol.~48, no.~1, p.~73, 1935.

\bibitem{piroli2020random}
L.~Piroli, C.~S{\"u}nderhauf, and X.-L. Qi, ``A random unitary circuit model
  for black hole evaporation,'' {\em Journal of High Energy Physics: JHEP},
  vol.~2020, no.~4, p.~63, 2020.

\bibitem{agarwal2019toy}
K.~Agarwal and N.~Bao, ``A toy model for decoherence in the black hole
  information problem,'' {\em arXiv preprint arXiv:1912.09491}, 2019.

\bibitem{akers2019simple}
C.~Akers, N.~Engelhardt, and D.~Harlow, ``Simple holographic models of black
  hole evaporation,'' {\em arXiv preprint arXiv:1910.00972}, 2019.

\bibitem{landsman2019verified}
K.~A. Landsman, C.~Figgatt, T.~Schuster, N.~M. Linke, B.~Yoshida, N.~Y. Yao,
  and C.~Monroe, ``Verified quantum information scrambling,'' {\em Nature},
  vol.~567, no.~7746, pp.~61--65, 2019.

\bibitem{yoshida2017efficient}
B.~Yoshida and A.~Kitaev, ``Efficient decoding for the hayden-preskill
  protocol,'' {\em arXiv preprint arXiv:1710.03363}, 2017.

\bibitem{blok2020quantum}
M.~Blok, V.~V. Ramasesh, T.~Schuster, K.~O'Brien, J.~M. Kreikebaum, D.~Dahlen,
  A.~Morvan, B.~Yoshida, N.~Y. Yao, and I.~Siddiqi, ``Quantum information
  scrambling in a superconducting qutrit processor,'' {\em arXiv preprint
  arXiv:2003.03307}, 2020.

\bibitem{banks1997m}
T.~Banks, W.~Fischler, S.~H. Shenker, and L.~Susskind, ``M theory as a matrix
  model: A conjecture,'' {\em Physical Review D}, vol.~55, no.~8, p.~5112,
  1997.

\bibitem{itzhaki1998supergravity}
N.~Itzhaki, J.~M. Maldacena, J.~Sonnenschein, and S.~Yankielowicz,
  ``Supergravity and the large n limit of theories with sixteen supercharges,''
  {\em Physical Review D}, vol.~58, no.~4, p.~046004, 1998.

\bibitem{gubser1998gauge}
S.~S. Gubser, I.~R. Klebanov, and A.~M. Polyakov, ``Gauge theory correlators
  from non-critical string theory,'' {\em Physics Letters B}, vol.~428,
  no.~1-2, pp.~105--114, 1998.

\bibitem{witten1998anti}
E.~Witten, ``Anti de sitter space and holography,'' {\em arXiv preprint
  hep-th/9802150}, 1998.

\bibitem{maldacena1999large}
J.~Maldacena, ``The large-n limit of superconformal field theories and
  supergravity,'' {\em International journal of theoretical physics}, vol.~38,
  no.~4, pp.~1113--1133, 1999.

\bibitem{aharonov1964time}
Y.~Aharonov, P.~G. Bergmann, and J.~L. Lebowitz, ``Time symmetry in the quantum
  process of measurement,'' {\em Physical Review}, vol.~134, no.~6B, p.~B1410,
  1964.

\bibitem{aharonov1988result}
Y.~Aharonov, D.~Z. Albert, and L.~Vaidman, ``How the result of a measurement of
  a component of the spin of a spin-1/2 particle can turn out to be 100,'' {\em
  Physical review letters}, vol.~60, no.~14, p.~1351, 1988.

\bibitem{duck1989sense}
I.~M. Duck, P.~M. Stevenson, and E.~C.~G. Sudarshan, ``The sense in which a"
  weak measurement" of a spin-$1/2$ particle's spin component yields a value
  100,'' {\em Physical Review D}, vol.~40, no.~6, p.~2112, 1989.

\bibitem{schwarzschild1916uber}
K.~Schwarzschild, ``Uber das gravitationsfeld eines massenpunktes nach der
  einstein'schen theorie,'' {\em Berlin. Sitzungsberichte}, vol.~18, 1916.

\bibitem{unruh1976notes}
W.~G. Unruh, ``Notes on black-hole evaporation,'' {\em Physical Review D},
  vol.~14, no.~4, p.~870, 1976.

\bibitem{jacobson2013black}
T.~Jacobson, ``Black holes and hawking radiation in spacetime and its
  analogues,'' in {\em Analogue Gravity Phenomenology}, pp.~1--29, Springer,
  2013.

\bibitem{guinea1988quantum}
F.~Guinea and G.~Sch{\"o}n, ``Quantum fluctuations in normal
  metal-superconductor and superconductor-normal metal-superconductor
  devices,'' {\em Physica B: Condensed Matter}, vol.~152, no.~1-2,
  pp.~165--171, 1988.

\bibitem{anderson1959theory}
P.~W. Anderson, ``Theory of dirty superconductors,'' {\em Journal of Physics
  and Chemistry of Solids}, vol.~11, no.~1-2, pp.~26--30, 1959.

\bibitem{bardeen1957theory}
J.~Bardeen, L.~N. Cooper, and J.~R. Schrieffer, ``Theory of
  superconductivity,'' {\em Physical review}, vol.~108, no.~5, p.~1175, 1957.

\bibitem{puspus2014entanglement}
X.~M. Puspus, K.~H. Villegas, and F.~N.~C. Paraan, ``Entanglement spectrum and
  number fluctuations in the spin-partitioned bcs ground state,'' {\em Physical
  Review B}, vol.~90, no.~15, p.~155123, 2014.

\bibitem{chakraborty2017black}
S.~Chakraborty and K.~Lochan, ``Black holes: Eliminating information or
  illuminating new physics?,'' {\em Universe}, vol.~3, no.~3, p.~55, 2017.

\bibitem{page1993average}
D.~N. Page, ``Average entropy of a subsystem,'' {\em Physical review letters},
  vol.~71, no.~9, p.~1291, 1993.

\bibitem{page1993information}
D.~N. Page, ``Information in black hole radiation,'' {\em Physical review
  letters}, vol.~71, no.~23, p.~3743, 1993.

\bibitem{sekino2008fast}
Y.~Sekino and L.~Susskind, ``Fast scramblers,'' {\em Journal of High Energy
  Physics}, vol.~2008, no.~10, p.~065, 2008.

\bibitem{susskind2011addendum}
L.~Susskind, ``Addendum to fast scramblers,'' {\em arXiv preprint
  arXiv:1101.6048}, 2011.

\bibitem{penington2019entanglement}
G.~Penington, ``Entanglement wedge reconstruction and the information
  paradox,'' {\em arXiv preprint arXiv:1905.08255}, 2019.

\bibitem{almheiri2019entropy}
A.~Almheiri, N.~Engelhardt, D.~Marolf, and H.~Maxfield, ``The entropy of bulk
  quantum fields and the entanglement wedge of an evaporating black hole,''
  {\em Journal of High Energy Physics}, vol.~2019, no.~12, p.~63, 2019.

\bibitem{czech2012gravity}
B.~Czech, J.~L. Karczmarek, F.~Nogueira, and M.~Van~Raamsdonk, ``The gravity
  dual of a density matrix,'' {\em Classical and Quantum Gravity}, vol.~29,
  no.~15, p.~155009, 2012.

\bibitem{headrick2014causality}
M.~Headrick, V.~E. Hubeny, A.~Lawrence, and M.~Rangamani, ``Causality \&
  holographic entanglement entropy,'' {\em Journal of High Energy Physics},
  vol.~2014, no.~12, p.~162, 2014.

\bibitem{wall2014maximin}
A.~C. Wall, ``Maximin surfaces, and the strong subadditivity of the covariant
  holographic entanglement entropy,'' {\em Classical and Quantum Gravity},
  vol.~31, no.~22, p.~225007, 2014.

\bibitem{rensink1967off}
M.~E. Rensink, ``Off-diagonal long-range order in the bcs theory,'' {\em Annals
  of Physics}, vol.~44, no.~1, pp.~105--111, 1967.

\bibitem{deutscher2002crossed}
G.~Deutscher, ``Crossed andreev reflections,'' {\em Journal of
  superconductivity}, vol.~15, no.~1, pp.~43--47, 2002.

\bibitem{bennett1997strengths}
C.~H. Bennett, E.~Bernstein, G.~Brassard, and U.~Vazirani, ``Strengths and
  weaknesses of quantum computing,'' {\em SIAM journal on Computing}, vol.~26,
  no.~5, pp.~1510--1523, 1997.

\bibitem{grover1996fast}
L.~K. Grover, ``A fast quantum mechanical algorithm for database search,'' in
  {\em Proceedings of the twenty-eighth annual ACM symposium on Theory of
  computing}, pp.~212--219, 1996.

\bibitem{harlow2013quantum}
D.~Harlow and P.~Hayden, ``Quantum computation vs. firewalls,'' {\em Journal of
  High Energy Physics}, vol.~2013, no.~6, p.~85, 2013.

\bibitem{dial2013charge}
O.~E. Dial, M.~D. Shulman, S.~P. Harvey, H.~Bluhm, V.~Umansky, and A.~Yacoby,
  ``Charge noise spectroscopy using coherent exchange oscillations in a
  singlet-triplet qubit,'' {\em Physical review letters}, vol.~110, no.~14,
  p.~146804, 2013.

\bibitem{watson2017atomically}
T.~F. Watson, B.~Weber, Y.-L. Hsueh, L.~C. Hollenberg, R.~Rahman, and M.~Y.
  Simmons, ``Atomically engineered electron spin lifetimes of 30 s in
  silicon,'' {\em Science advances}, vol.~3, no.~3, p.~e1602811, 2017.

\bibitem{baart2016single}
T.~A. Baart, M.~Shafiei, T.~Fujita, C.~Reichl, W.~Wegscheider, and L.~M.~K.
  Vandersypen, ``Single-spin ccd,'' {\em Nature nanotechnology}, vol.~11,
  no.~4, p.~330, 2016.

\bibitem{tyryshkin2012electron}
A.~M. Tyryshkin, S.~Tojo, J.~J. Morton, H.~Riemann, N.~V. Abrosimov, P.~Becker,
  H.-J. Pohl, T.~Schenkel, M.~L. Thewalt, K.~M. Itoh, {\em et~al.}, ``Electron
  spin coherence exceeding seconds in high-purity silicon,'' {\em Nature
  materials}, vol.~11, no.~2, pp.~143--147, 2012.

\bibitem{hanson2007spins}
R.~Hanson, L.~P. Kouwenhoven, J.~R. Petta, S.~Tarucha, and L.~M.~K.
  Vandersypen, ``Spins in few-electron quantum dots,'' {\em Reviews of modern
  physics}, vol.~79, no.~4, p.~1217, 2007.

\bibitem{scarlino2014spin}
P.~Scarlino, E.~Kawakami, P.~Stano, M.~Shafiei, C.~Reichl, W.~Wegscheider, and
  L.~M.~K. Vandersypen, ``Spin-relaxation anisotropy in a gaas quantum dot,''
  {\em Physical review letters}, vol.~113, no.~25, p.~256802, 2014.

\bibitem{maldacena2013cool}
J.~Maldacena and L.~Susskind, ``Cool horizons for entangled black holes,'' {\em
  Fortschritte der Physik}, vol.~61, no.~9, pp.~781--811, 2013.

\end{thebibliography}
\bibliographystyle{ieeetr}
		   \end{document}